\documentclass[aps,prl,twocolumn,showpacs,superscriptaddress,groupedaddress]{revtex4}  
\usepackage{blindtext}
\usepackage{natbib}
\usepackage{graphicx}
\usepackage{amssymb, amsmath}
\usepackage{hyperref}
\usepackage{color}
\usepackage[normalem]{ulem}

\begin{document}

\title{Constraints on Minute-Scale Transient Astrophysical Neutrino Sources}
\affiliation{III. Physikalisches Institut, RWTH Aachen University, D-52056 Aachen, Germany}
\affiliation{Department of Physics, University of Adelaide, Adelaide, 5005, Australia}
\affiliation{Department~of Physics and Astronomy, University of Alaska Anchorage, 3211 Providence Dr., Anchorage, AK 99508, USA}
\affiliation{Department~of Physics, University of Texas at Arlington, 502 Yates St., Science Hall Rm 108, Box 19059, Arlington, TX 76019, USA}
\affiliation{CTSPS, Clark-Atlanta University, Atlanta, GA 30314, USA}
\affiliation{School of Physics and Center for Relativistic Astrophysics, Georgia Institute of Technology, Atlanta, GA 30332, USA}
\affiliation{Department~of Physics, Southern University, Baton Rouge, LA 70813, USA}
\affiliation{Department~of Physics, University of California, Berkeley, CA 94720, USA}
\affiliation{Lawrence Berkeley National Laboratory, Berkeley, CA 94720, USA}
\affiliation{Institut f\"ur Physik, Humboldt-Universit\"at zu Berlin, D-12489 Berlin, Germany}
\affiliation{Fakult\"at f\"ur Physik \& Astronomie, Ruhr-Universit\"at Bochum, D-44780 Bochum, Germany}
\affiliation{Universit\'e Libre de Bruxelles, Science Faculty CP230, B-1050 Brussels, Belgium}
\affiliation{Vrije Universiteit Brussel (VUB), Dienst ELEM, B-1050 Brussels, Belgium}
\affiliation{Department~of Physics, Massachusetts Institute of Technology, Cambridge, MA 02139, USA}
\affiliation{Department of Physics and Institute for Global Prominent Research, Chiba University, Chiba 263-8522, Japan}
\affiliation{Department of Physics and Astronomy, University of Canterbury, Private Bag 4800, Christchurch, New Zealand}
\affiliation{Department of Physics, University of Maryland, College Park, MD 20742, USA}
\affiliation{Department of Physics and Center for Cosmology and Astro-Particle Physics, Ohio State University, Columbus, OH 43210, USA}
\affiliation{Department of Astronomy, Ohio State University, Columbus, OH 43210, USA}
\affiliation{Niels Bohr Institute, University of Copenhagen, DK-2100 Copenhagen, Denmark}
\affiliation{Department of Physics, TU Dortmund University, D-44221 Dortmund, Germany}
\affiliation{Department of Physics and Astronomy, Michigan State University, East Lansing, MI 48824, USA}
\affiliation{Department of Physics, University of Alberta, Edmonton, Alberta, Canada T6G 2E1}
\affiliation{Erlangen Centre for Astroparticle Physics, Friedrich-Alexander-Universit\"at Erlangen-N\"urnberg, D-91058 Erlangen, Germany}
\affiliation{D\'epartement de physique nucl\'eaire et corpusculaire, Universit\'e de Gen\`eve, CH-1211 Gen\`eve, Switzerland}
\affiliation{Department of Physics and Astronomy, University of Gent, B-9000 Gent, Belgium}
\affiliation{Department of Physics and Astronomy, University of California, Irvine, CA 92697, USA}
\affiliation{Department of Physics and Astronomy, University of Kansas, Lawrence, KS 66045, USA}
\affiliation{Department of Physics and Astronomy, University of Leicester, Leicester LE1 7RH, UK}
\affiliation{SNOLAB, 1039 Regional Road 24, Creighton Mine 9, Lively, ON, Canada P3Y 1N2}
\affiliation{Department of Physics and Astronomy, UCLA, Los Angeles, CA 90095, USA}
\affiliation{Department of Astronomy, University of Wisconsin, Madison, WI 53706, USA}
\affiliation{Department of Physics and Wisconsin IceCube Particle Astrophysics Center, University of Wisconsin, Madison, WI 53706, USA}
\affiliation{Institute of Physics, University of Mainz, Staudinger Weg 7, D-55099 Mainz, Germany}
\affiliation{Department of Physics, Marquette University, Milwaukee, WI, 53201, USA}
\affiliation{Physik-department, Technische Universit\"at M\"unchen, D-85748 Garching, Germany}
\affiliation{Institut f\"ur Kernphysik, Westf\"alische Wilhelms-Universit\"at M\"unster, D-48149 M\"unster, Germany}
\affiliation{Bartol Research Institute and Department of Physics and Astronomy, University of Delaware, Newark, DE 19716, USA}
\affiliation{Department of Physics, Yale University, New Haven, CT 06520, USA}
\affiliation{Department of Physics, University of Oxford, 1 Keble Road, Oxford OX1 3NP, UK}
\affiliation{Department of Physics, Drexel University, 3141 Chestnut Street, Philadelphia, PA 19104, USA}
\affiliation{Physics Department, South Dakota School of Mines and Technology, Rapid City, SD 57701, USA}
\affiliation{Department of Physics, University of Wisconsin, River Falls, WI 54022, USA}
\affiliation{Department of Physics and Astronomy, University of Rochester, Rochester, NY 14627, USA}
\affiliation{Oskar Klein Centre and Department of Physics, Stockholm University, SE-10691 Stockholm, Sweden}
\affiliation{Department of Physics and Astronomy, Stony Brook University, Stony Brook, NY 11794-3800, USA}
\affiliation{Department of Physics, Sungkyunkwan University, Suwon 440-746, Korea}
\affiliation{Department of Physics and Astronomy, University of Alabama, Tuscaloosa, AL 35487, USA}
\affiliation{Department of Astronomy and Astrophysics, Pennsylvania State University, University Park, PA 16802, USA}
\affiliation{Department of Physics, Pennsylvania State University, University Park, PA 16802, USA}
\affiliation{Department of Physics and Astronomy, Uppsala University, Box 516, S-75120 Uppsala, Sweden}
\affiliation{Department of Physics, University of Wuppertal, D-42119 Wuppertal, Germany}
\affiliation{DESY, D-15738 Zeuthen, Germany}

\author{M.~G.~Aartsen}
\affiliation{Department of Physics and Astronomy, University of Canterbury, Private Bag 4800, Christchurch, New Zealand}
\author{M.~Ackermann}
\affiliation{DESY, D-15738 Zeuthen, Germany}
\author{J.~Adams}
\affiliation{Department of Physics and Astronomy, University of Canterbury, Private Bag 4800, Christchurch, New Zealand}
\author{J.~A.~Aguilar}
\affiliation{Universit\'e Libre de Bruxelles, Science Faculty CP230, B-1050 Brussels, Belgium}
\author{M.~Ahlers}
\affiliation{Niels Bohr Institute, University of Copenhagen, DK-2100 Copenhagen, Denmark}
\author{M.~Ahrens}
\affiliation{Oskar Klein Centre and Department of Physics, Stockholm University, SE-10691 Stockholm, Sweden}
\author{I.~Al~Samarai}
\affiliation{D\'epartement de physique nucl\'eaire et corpusculaire, Universit\'e de Gen\`eve, CH-1211 Gen\`eve, Switzerland}
\author{D.~Altmann}
\affiliation{Erlangen Centre for Astroparticle Physics, Friedrich-Alexander-Universit\"at Erlangen-N\"urnberg, D-91058 Erlangen, Germany}
\author{K.~Andeen}
\affiliation{Department of Physics, Marquette University, Milwaukee, WI, 53201, USA}
\author{T.~Anderson}
\affiliation{Department of Physics, Pennsylvania State University, University Park, PA 16802, USA}
% ...

\date{\today}

\begin{abstract}
High-energy neutrino emission has been predicted for several short-lived astrophysical transients including gamma-ray bursts (GRBs), core-collapse supernovae with choked jets and neutron star mergers. IceCube's optical and x-ray follow-up program searches for such transient sources by looking for two or more muon neutrino candidates in directional coincidence and arriving within 100\,s. The measured rate of neutrino alerts is consistent with the expected rate of chance coincidences of atmospheric background events and no likely electromagnetic counterparts have been identified in \emph{Swift} follow-up observations. Here, we calculate generic bounds on the neutrino flux of short-lived transient sources. Assuming an $E^{-2.5}$ neutrino spectrum, we find that the neutrino flux of rare sources, like long gamma-ray bursts, is constrained to $<5$\% of the detected astrophysical flux and the energy released in neutrinos (100\,GeV to 10\,PeV) by a median bright GRB-like source is $<10^{52.5}$\,erg. For a harder $E^{-2.13}$ neutrino spectrum up to $30$\% of the flux could be produced by GRBs and the allowed median source energy is $< 10^{52}$\,erg. A hypothetical population of transient sources has to be more common than $10^{-5}\text{\,Mpc}^{-3}\text{\,yr}^{-1}$ ($5\times10^{-8}\,\text{Mpc}^{-3}\,\text{yr}^{-1}$ for the $E^{-2.13}$ spectrum) to account for the complete astrophysical neutrino flux.

\end{abstract}

\maketitle


\section{Introduction}
An astrophysical neutrino flux at high energies (from $\sim$10\,TeV to a few PeV) was discovered by the IceCube neutrino observatory \cite{icecube2013b, icecube2014b, icecube2016c}. The neutrino arrival directions are largely isotropic suggesting a predominantly extragalactic origin. Possible sources include long gamma-ray bursts (GRBs) \cite{waxman1997, guetta2004, meszaros2006, baerwald2014}, core-collapse supernovae (\mbox{CCSNe}) with choked jets \cite{fraija2014, senno2016, tamborra2016} binary neutron star mergers \cite{kimura2017, biehl2018} and active galactic nuclei (AGNs) \cite{stecker1991, sironi2011, essey2010, kalashev2013, murase2014b} (see e.g. Ref. \cite{murase2014}, for a more extensive list).
While several neutrino events have been associated with a blazar \cite{icecube2018, icecube2018b}, blazars likely cannot account for the complete astrophysical flux \cite{icecube2017d}. The absence of luminous neutrino point sources \cite{icecube2016, icecube2016c, reimann2017} implies that the observed flux can only be emitted by a class of sufficiently numerous sources \cite{lipari2008,ahlers2014,murase2016,glauch2017}.


The IceCube detector is deployed in the glacial ice at the geographical South Pole at depths between 1450 to 2450\,m and comprises a volume of $1\text{\,km}^3$ \cite{icecube2016b}. It detects neutrino events with energies between 100\,GeV and a few PeV. If a secondary muon is produced in a neutrino interaction, its tracklike signature allows us to resolve the neutrino direction to $\sim1^\circ$ \cite{icecube2016}. IceCube has a dedicated optical and x-ray follow-up program which is triggered by two or more tracklike events detected within $<100\,\text{s}$ that are consistent with a point source origin \cite{icecube2012b, icecube2015, icecube2017}. Except for AGNs, the above-mentioned source classes are all expected to produce such short neutrino bursts as they are powered by central engines which are typically active for few to about $100\,\text{s}$.
To look for a potential electromagnetic counterpart, follow-up observations for the least backgroundlike alerts are obtained with the X-Ray Telescope (XRT \cite{burrows2005}) on board the Neil Gehrels \emph{Swift} observatory, the 48-inch telescope of the Palomar Transient Factory (PTF \cite{law2009, rau2009}; until Feb. 2017), and the Robotic Optical Transient Search Experiment (ROTSE \cite{akerlof2003}; until Nov. 2015).

So far, no optical or x-ray transient sources have been positively associated with any of the neutrino multiplets \cite{evans2015, icecube2015, icecube2017}. As the alert rates are consistent with the background-only hypothesis, we find that strong constraints on the existence of short-lived transient populations can be derived from the IceCube data alone.

\section{Detected neutrino alerts}

IceCube's optical and x-ray follow-up program was established in Dec. 2008 to search for short-lived transient neutrino sources and here we present results from the first five years of operation with the complete detector (Sept. 2011 -- May 2016).

For the follow-up program we select tracklike events, called neutrino candidates, from the northern sky (for a detailed description of the event selection see Ref.~\cite{icecube2016d}) which are detected at a rate of about $3$\,mHz. To suppress the dominating background of atmospheric neutrino and muon events we search for two or more neutrino candidates with a temporal separation of less than 100\,s and an angular separation of less than $3.5^\circ$. Doublets are alerts consisting of two neutrino candidates, while we call alerts with three or more candidates \emph{multiplets}.

Within the live time of 1648.1 days we selected in total 460\,438 neutrino candidates. The selected data consist of about $\sim80$\% atmospheric neutrinos, $\sim20$\% misreconstructed atmospheric muons from the southern sky \citep{voge2016}, and less than 1\% astrophysical neutrinos depending on the assumed spectral shape of the astrophysical neutrino flux.

Alerts can also be produced by chance coincidences of background events and we calculate the rate of background alerts by randomizing the detection times of events, as described in Ref.~\cite{icecube2017}. The expected background is 312.7 doublets, 0.341 triplets and only $5\times10^{-4}$ quadruplets within the analyzed live time. We have observed 338 neutrino doublets 
and one neutrino triplet \cite{icecube2017} 
(see Supplemental Material for more detail on the alerts \footnote{\label{supplemental}See Supplemental Material at the end of this document}). The resulting 90\% upper limit \cite{feldman1998} on the number of astrophysical doublets is $<56$, while the limit on the expected number of astrophysical triplets is $<4.0$ within the analyzed live time. We find that the triplet rate provides stronger constraints on the neutrino flux of transient source populations.

The significance of doublet alerts is quantified as described in Ref.~\cite{icecube2015}, but all alerts were consistent with being chance coincidences of atmospheric events. The two most significant alerts were studied in great detail \cite{icecube2015, icecube2017} and no likely electromagnetic counterpart was detected. \emph{Swift} XRT follow-up observations have been obtained for $25$ alerts and no sources were identified above a predefined threshold (see Ref.~\cite{evans2015}). 

The alert rates, doublet significances and \emph{Swift} XRT follow-up observations hence do not provide evidence for the existence of a population of short-lived transient sources.
In the following we therefore do not make use of the collected follow-up observations, but use the low rate of alerts with three or more neutrino candidates to calculate generic constraints on the neutrino emission of short-lived transient populations like GRBs and CCSNe.

\section{Simulating transient source populations}

The low rate of detected neutrino multiplets allows us to calculate limits on the neutrino flux of a population of transient sources with durations up to 100\,s. For this purpose we simulate two types of transient source populations whose properties are chosen such that they are similar to long GRBs and CCSNe with a choked jet. The impact of the different assumptions on the results is summarized in Table 3 of the Supplemental Material \cite{endnote58}.

The redshift distributions for GRBs and CCSNe are taken from Refs.~\cite{wanderman2010} and \cite{madau2014} respectively. The distribution for CCSNe peaks at a lower redshift of $z\sim2$ compared to the one for GRBs which peaks at $z\sim3$. We simulate sources in the northern sky up to a redshift of $z=8$ and use the cosmological parameters from Ref.~\cite{planck2016}. Sources located at $z>4$ only contribute $1$\% ($5$\%) of the events for the CCSN-like (GRB-like) population and hence only have a small effect on the results.

The distribution of GRB peak luminosities is relatively broad, spanning at least 4 orders of magnitude \cite{wanderman2010}. We assume that the neutrino peak luminosities of GRBs follow the distribution measured in gamma rays. The population of CCSNe does not show as large luminosity fluctuations at the optical wavelengths \cite{richardson2014} and we assume a narrow log-normal distribution with a width of $0.4$ in log-10 space corresponding to fluctuations of one astronomical magnitude. The fluctuations assumed for the GRB-like population are larger by a factor of 300. Ultimately the neutrino luminosity functions of both populations are unknown, and the two different scenarios allow us to quantify their influence on the detection probability.

Transient durations in the source rest frame are drawn from a log-normal distribution centered around $11.2$\,s with a width of $0.58$ in log-10 space, which approximately reproduces the duration distribution of long GRBs measured at Earth \footnote{The durations of long GRBs from the \emph{Swift} catalog are taken from \url{http://swift.gsfc.nasa.gov/archive/grb_table/}.}. We hence assume that the duration of the neutrino and gamma-ray emission is similar. CCSNe with choked jets have not yet been observed, but we chose to use the same duration distribution. We assume that the transient source instantaneously rises to its peak luminosity and then decays exponentially according to its simulated duration.
The number of multiplet alerts does not depend on the shape of the light curve as long as the neutrinos arrive within $100$\,s.

The neutrino emission of each source is assumed to follow a power-law spectrum similar to the detected astrophysical neutrino flux
\begin{equation}
\label{eq:spec}
\phi(E) = \phi_0 \times (E/\text{GeV})^{-\gamma} \quad .
\end{equation}

To account for the uncertainty on the measured neutrino flux, we use two different spectral shapes: a hard spectrum with $\gamma=2.13$ and $\phi_0 = 4.0\times10^{-8}\text{\,GeV}^{-1}\text{cm}^{-2} \text{s}^{-1} \text{sr}^{-1}$ and a soft spectrum with $\gamma=2.5$ and $\phi_0 = 7.1\times10^{-6}\text{\,GeV}^{-1}\text{cm}^{-2} \text{s}^{-1} \text{sr}^{-1}$.
The normalization $\phi_0$ is per neutrino flavor and includes both neutrinos and antineutrinos.
The soft spectrum has been measured in a global fit extending down to an energy of $10$\,TeV \cite{icecube2015c} while the hard $E^{-2.13}$ spectrum was found in an analysis restricted to tracklike events from the northern sky with energies $\geqslant\!100$\,TeV~\cite{icecube2016c}. 

The sensitivity of the follow-up program is evaluated using simulated IceCube neutrino events accounting for the detector acceptance and the effects of high-energy neutrino absorption in Earth's core. During the data-taking period, data selection methods and reconstructions have been steadily improved. We account for these changes in our simulations.

\begin{figure}[tb]
\includegraphics[width=\columnwidth]{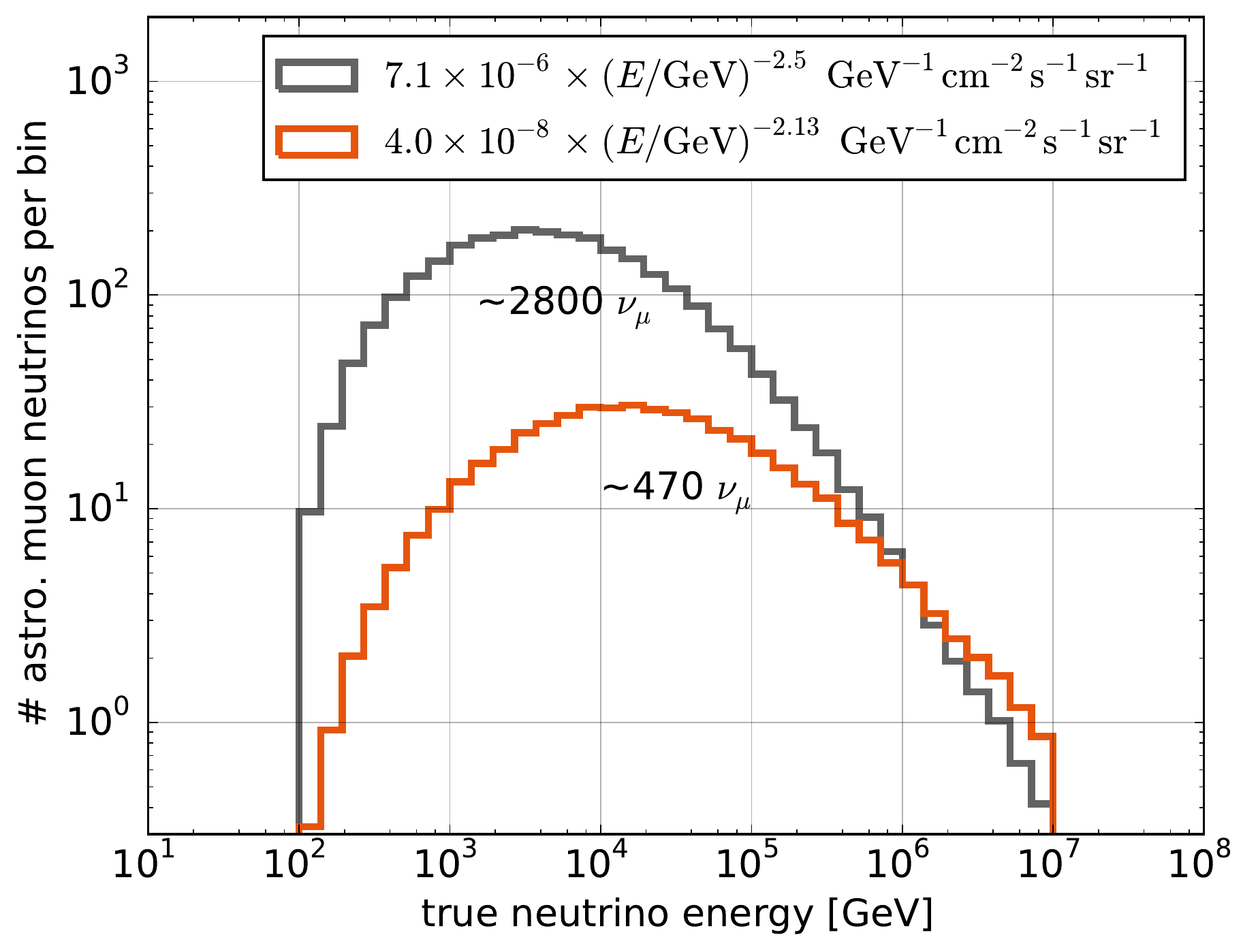}
\caption{\label{fig:neutrino_spectra} Expected number of astrophysical neutrinos passing the event selection of the follow-up program within 1648.1 day live time. Two different fits to the measured flux are adopted (see Eq.~\ref{eq:spec}). The reconstructed energy can be much lower than the true neutrino energy shown here, since most tracklike events are not contained within the instrumented volume.}
\end{figure}

The energy distributions of the events which pass all selection cuts are shown in Fig.~\ref{fig:neutrino_spectra}. The total expected number of astrophysical neutrino track events within the livetime of 1648.1 days is about $470$ and $2800$ $\nu_\mu$ for the $E^{-2.13}$ and $E^{-2.5}$ spectrum respectively (see Table 2 in the Supplemental Material \cite{endnote58} for more details). Here we extrapolate the power-law neutrino flux down to 100\,GeV. Such a spectrum is expected if the neutrinos are produced in $pp$ interactions; however for $p\gamma$ interactions there would be a low-energy cutoff \cite{murase2016}. Above the threshold of 10\,TeV, where the astrophysical flux is constrained by data \cite{icecube2014c}, we expect about $280$ or $910$ $\nu_\mu$, respectively.

\section{Generic constraints}

The simulated source populations are used to infer limits on the neutrino emission of short transient sources. We vary both the rate of sources, and the neutrino flux emitted by the complete population, to rule out scenarios that produce more than one detected neutrino multiplet within the analyzed live time at 90\% confidence level.

\begin{table}[b]
\caption{\label{tab:expected_multiplicities} Expected number of alerts from simulated source populations and 90\% upper limits on their neutrino emission. The limits were calculated based on the observation of only one neutrino triplet within the analyzed live time.}
\begin{ruledtabular}
\begin{tabular}{l c c c c}
Population & \multicolumn{2}{c}{\textbf{long GRBs}}      & \multicolumn{2}{c}{\textbf{1\% of CCSNe}} \\
Spectral shape	& $E^{-2.13}$	& $E^{-2.5}$        & $E^{-2.13}$ 		& $E^{-2.5}$\\
\hline
Rate [Mpc$^{-3}$\,yr$^{-1}$]	&  \multicolumn{2}{c}{$4.2\times10^{-10}$} & \multicolumn{2}{c}{$6.8\times10^{-7}$} \\
No. sources$^a$	        	&  \multicolumn{2}{c}{7200} 		& \multicolumn{2}{c}{$5.9\times10^6$} \\
\hline
\multicolumn{3}{l}{\textbf{Expected no. of alerts:}\,$^b$} & & \\
No. singlets ($1\nu_\mu$)     & 0 (143)       & 0 (339)         	& 0 (450)             	& 0 (2470) \\
No. doublets ($2\nu_\mu$)       	& 16 (26)     	& 58 (92)    		& 2.3 (4.0)     	& 33 (60)\\
No. multiplets ($\geqslant3\nu_\mu$)     & 22 (28)    	& 119 (144)   		& 1.1 (1.5)     	& 19 (26)\\
\hline
\multicolumn{3}{l}{\textbf{Resulting limits:}\,$^c$} & & \\
Frac. of diffuse flux  		& $<$30\% 	    & $<$5\% 		    & $<$250\% 		    & $<$40\%\\
Source $\nu$ energy [erg]  	&$<\!10^{52}$ & $<\!10^{52.5}$	& $<\!10^{50.5}$\,$^d$	& $<\!10^{50.8}$\\
\end{tabular}
\end{ruledtabular}
\begin{flushleft}
\small{
$^a$ Number of transients in the Northern sky within $z\leqslant8$ within the live time of 1648.1 days.\\
$^b$ Expected number of signal doublets and multiplets if the respective population accounts for 100\% of the astrophysical neutrino flux. The numbers in parentheses do not include losses due to our cuts (two events within $<3.5^\circ$ and 100\,s). The total number of expect events is $\sim470$ for an $E^{-2.13}$ spectrum and $\sim2800$ for an $E^{-2.5}$ spectrum.\\
$^c$ 90\% c.l. upper limits on the neutrino emission ($100$\,GeV to $10$\,PeV; flavor equipartition) based on the detection of only one multiplet.\\
$^d$ The detected astrophysical flux yields a more constraining limit on the energy emitted in neutrinos of $<\!10^{50.1}$\,erg.}
\end{flushleft}
\end{table}

While the source rate is a free parameter in the final result, we discuss in addition the results for two measured transient rates in more detail: In the first example we constrain the neutrino emission of a GRB-like population while in the second one we assume that 1\% of all CCSNe contribute to the astrophysical neutrino flux (e.g., because they contain choked jets pointed towards Earth; see also Refs.~\cite{soderberg2006, sobacchi2017, denton2017}). The local rates of GRBs and CCSNe are taken from Refs.~\cite{lien2014, lien2015} and~\cite{strolger2015}, respectively. They allow us to convert between the local source rate and the number of transients (see Table~\ref{tab:expected_multiplicities}).

We then vary the neutrino flux of the source populations and calculate the expected number of detected neutrino events for each source. This depends on the source redshift, peak luminosity, transient duration and zenith direction. We use a Poisson distribution to calculate how likely it is that one, two, or more than two neutrinos are detected from a source (shown in parentheses in Table~\ref{tab:expected_multiplicities}).

The probability that the reconstructed directions of two neutrinos from the same source are separated by more than $3.5^\circ$ depends strongly on the neutrino energies and zenith direction with a median probability of 27\% for the $E^{-2.5}$ spectrum. Additional losses occur when the neutrinos arrive more than 100\,s apart, which happens for $9\%$ of the sources for the assumed duration and redshift distribution. Assuming that the population produces the entire astrophysical neutrino flux, the expected number of astrophysical doublet and multiplet alerts is shown in the middle part of Table~\ref{tab:expected_multiplicities}. Sources with a single detected event cannot produce an alert.

Using the Feldman Cousins method \citep{feldman1998}, we rule out scenarios in which the detection of more than one multiplet from signal or background (0.341 chance coincidences) is expected with 90\% probability. We find that the expected number of astrophysical multiplets is $<4.0$ within the analyzed live time. We calculate limits on the population's neutrino emission and on the energy that the median source in the population can release in neutrinos in the energy range from 100\,GeV to 10\,PeV in the source rest frame.

Systematic errors on IceCube's sensitivity are dominated by the uncertainty on the optical efficiency of the detector and scattering and absorption in the ice. To quantify these uncertainties, we repeat the analysis with the efficiency reduced by 10\% and ice absorption increased by 10\%. Because of the lower number of detected neutrino events and the worse angular resolution, the number of multiplets decreases by 17\% (14\%) for the $E^{-2.5}$ ($E^{-2.13}$) spectrum.

\begin{figure}[tb]
\includegraphics[width=\columnwidth]{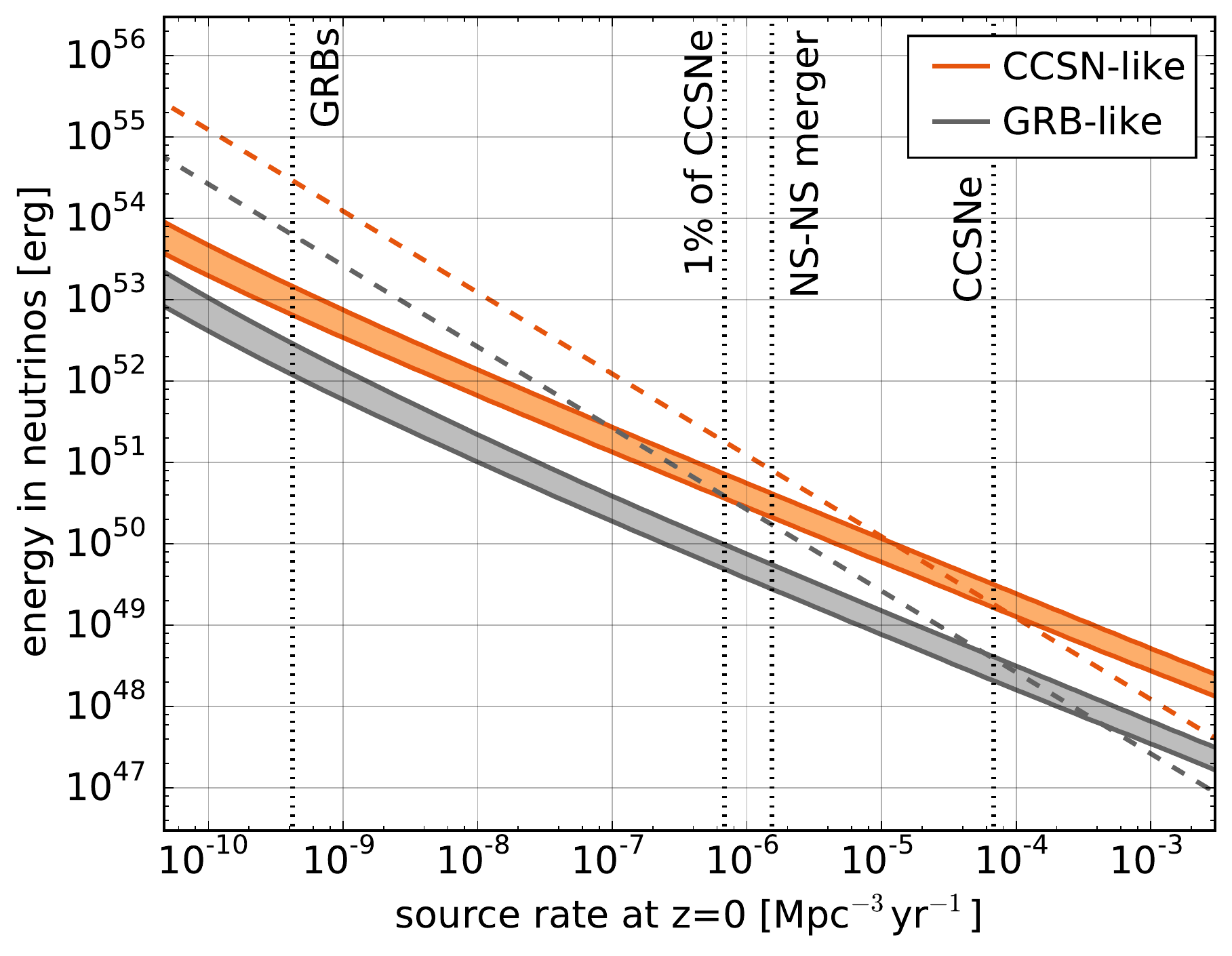}
\caption{\label{fig:exclusion} Limits on the median source energy (90\% c. l.) emitted in neutrinos between 100\,GeV and 10\,PeV within 100\,s. The area above the bands is excluded for CCSN-like (orange) and GRB-like (gray) populations respectively. The upper edge of the limit corresponds to an $E^{-2.5}$ neutrino spectrum and the lower one to an $E^{-2.13}$ spectrum. The diagonal dashed lines show which source energy accounts for 100\% of the astrophysical flux for an $E^{-2.5}$ spectrum. For the $E^{-2.13}$ spectrum, the complete flux is produced by 13 times fainter sources (lines not shown). The rate of long GRBs, NS-NS mergers and CCSNe is indicated. Beaming is included for long GRBs, but not for NS-NS mergers or CCSNe due to the unknown jet opening angles. The figure shows the limit on the median transient energy and the average energy is a factor of 3.8 (18) larger for the CCSN-like (GRB-like) population.}
\end{figure}

Figure~\ref{fig:exclusion} shows the upper limits, including systematic errors, on the median source energy for the GRB-like and SN-like source populations. The diagonal dashed lines indicate the median transient energy which would produce the complete detected flux. The corresponding lines for the harder $E^{-2.13}$ spectrum are a factor of 13 lower due to the extrapolation to lower energies (compare Fig.~\ref{fig:neutrino_spectra}). The ratio between the limits and the respective broken lines depicts the fraction of the detected astrophysical flux that a population with a given rate can at most produce (also given in the second last row of Table~\ref{tab:expected_multiplicities}). For populations consisting of many faint sources these lines provide more constraining limits, because only few multiplets are expected.

The study was repeated using only events with energies above 10\,TeV where the astrophysical flux has been measured. Without the extrapolation to 100\,GeV both neutrino spectra yield similar results (compare also Fig.~\ref{fig:neutrino_spectra}). The limit for the smaller energy range (shown in Fig.~1 in the Supplemental Material~\cite{endnote58}) is a factor of $\sim1.5$ lower compared to the lower edge of the bands shown in Fig.~\ref{fig:exclusion}, but corresponds to a larger fraction of the astrophysical neutrino flux.

The typical distance of a transient source that produces a neutrino multiplet depends on the source luminosity and on the source rate of the population, and is large for most considered rates (e.g. a median distance of $100\,\text{Mpc}$ for 1\% of the CCSN rate and the $E^{-2.13}$ neutrino spectrum). Only for the CCSN rate does the median distance decrease to $\sim10\,\text{Mpc}$, such that local inhomogeneities in the Universe might affect the multiplet rate \cite{tikhonov2009}.

As shown in Fig.~\ref{fig:exclusion} and Table~\ref{tab:expected_multiplicities}, we can constrain the neutrino emission from a GRB-like population to 5\% of the astrophysical flux adopting the $E^{-2.5}$ neutrino spectrum and to 30\% for the $E^{-2.13}$ spectrum. More frequent sources, such as NS-NS mergers \cite{ligo2017} or CCSNe, can account for much or all of the astrophysical neutrino flux. However, the rates shown for those two source classes do not include a beaming factor. If the neutrino emission is collimated in a jet the rate of observable transients would be reduced.

CCSN-like populations can only account for the complete astrophysical flux if their rate is larger than $10^{-5} \text{\,Mpc}^{-3}\text{\,yr}^{-1}$ ($5\times 10^{-8}\text{\,Mpc}^{-3}\text{\,yr}^{-1}$) for an $E^{-2.5}$ ($E^{-2.13}$) spectrum. We can hence exclude rare transients with less than 15\% (0.07\%) of the CCSN rate \cite{strolger2015} producing the entire astrophysical neutrino flux.

\section{Conclusion}

IceCube's optical and x-ray follow-up program triggers observations when multiple muon neutrino candidates are detected within 100\,s and are directionally consistent with a common source origin. The observed alert rates can be explained by background and no likely neutrino source has been identified. Extrapolating the detected astrophysical neutrino flux to 100\,GeV, we expect the detection of 470 to 2800 astrophysical muon neutrino events within the data collected over 1648.1 days. Based on the low rate of detected neutrino multiplets we calculate limits on the neutrino flux for two classes of short transient sources similar to GRBs and CCSNe with choked jets.

We find that a transient source population similar to long GRBs can at most account for 5\% (30\%) of the astrophysical neutrino flux for a neutrino spectrum of $E^{-2.5}$ ($E^{-2.13}$; see Fig.~\ref{fig:exclusion}). This corresponds to a limit on the energy emitted in neutrinos within 100\,s of $<10^{52.5}$\,erg ($<10^{52}$\,erg). Fewer neutrino multiplets are expected if the neutrino flux is emitted by a larger number of faint transients. A CCSN-like population can account for the complete flux if its rate at $z=0$ is larger than $10^{-5}\text{\,Mpc}^{-1}\text{\,yr}^{-1}$ ($5\times10^{-8}\text{\,Mpc}^{-1}\text{\,yr}^{-1}$).

The derived limits are valid for transient sources with durations up to 100\,s which follow the star formation rate or GRB redshift distribution. Dedicated searches for the neutrino emission from GRBs and CCSNe provide stronger constraints \cite{icecube2017b, stasik2018, icecube2019}. However, the limits derived here are more general: They are solely based on neutrino detections and therefore also apply to sources that are not detected in electromagnetic radiation or that exhibit a time delay between the neutrino and electromagnetic signal.
For binary neutron star mergers, the optimistic extended emission scenario in Ref.~\cite{kimura2017} would yield $\sim2$ detected neutrino multiplets within the analyzed live time and is hence within reach of the follow-up program. Different models \cite{kimura2017, antares2017, biehl2018}, however, predict source energies that are several orders of magnitude below the calculated limit.

The obtained limits strongly depend on the number of detected astrophysical neutrinos which is determined by the event selection, the assumed neutrino spectrum and the considered energy range. This is the likely cause for the different limits found in literature~\cite{ahlers2014, murase2016}. Contrary to previous analyses, our results are based on the full simulation of the IceCube detector including energy and directional dependent sensitivity and resolution, live time, event selection and alert generation. Our search for transient neutrino sources is ongoing \cite{icecube2016d} and real-time multiwavelength follow-up observations extend our sensitivity to sources which cannot be detected and identified by IceCube alone.

\begin{acknowledgments}

The authors gratefully acknowledge the support from the following agencies and institutions: 
USA -- U.S. National Science Foundation-Office of Polar Programs,
U.S. National Science Foundation-Physics Division,
Wisconsin Alumni Research Foundation,
Center for High Throughput Computing (CHTC) at the University of Wisconsin-Madison,
Open Science Grid (OSG),
Extreme Science and Engineering Discovery Environment (XSEDE),
U.S. Department of Energy-National Energy Research Scientific Computing Center,
Particle astrophysics research computing center at the University of Maryland,
Institute for Cyber-Enabled Research at Michigan State University,
and Astroparticle physics computational facility at Marquette University;
Belgium -- Funds for Scientific Research (FRS-FNRS and FWO),
FWO Odysseus and Big Science programmes,
and Belgian Federal Science Policy Office (Belspo);
Germany -- Bundesministerium f\"ur Bildung und Forschung (BMBF),
Deutsche Forschungsgemeinschaft (DFG),
Helmholtz Alliance for Astroparticle Physics (HAP),
Initiative and Networking Fund of the Helmholtz Association,
Deutsches Elektronen Synchrotron (DESY),
and High Performance Computing cluster of the RWTH Aachen;
Sweden -- Swedish Research Council,
Swedish Polar Research Secretariat,
Swedish National Infrastructure for Computing (SNIC),
and Knut and Alice Wallenberg Foundation;
Australia -- Australian Research Council;
Canada -- Natural Sciences and Engineering Research Council of Canada,
Calcul Qu\'ebec, Compute Ontario, Canada Foundation for Innovation, WestGrid, and Compute Canada;
Denmark -- Villum Fonden, Danish National Research Foundation (DNRF);
New Zealand -- Marsden Fund;
Japan -- Japan Society for Promotion of Science (JSPS)
and Institute for Global Prominent Research (IGPR) of Chiba University;
Korea -- National Research Foundation of Korea (NRF);
Switzerland -- Swiss National Science Foundation (SNSF).

This work made use of data supplied by the UK \emph{Swift} Science Data Centre at the University of Leicester. Funding for the Swift project in the UK is provided by the UK Space Agency.

We acknowledge the work of Andreas Homeier who contributed to the development of this analysis. 

\end{acknowledgments}


\bibliographystyle{apsrev} 
\bibliography{references}

\clearpage

 \begin{table*}[p]
\caption{\label{tab:events}Number of detected alerts compared to the expected number of alerts from background.}
\begin{ruledtabular}
\begin{tabular}{l c c c c c c c c c c}
& livetime & \multicolumn{2}{c}{Doublets} & \multicolumn{2}{c}{\emph{Swift} Doublets$^a$} & \multicolumn{2}{c}{Multiplets$^b$} & events & \multicolumn{2}{c}{astro. events$^c$}\\
Season & [days] & Det. & Exp.       & Det. & Exp. & Det. & Exp. &       & $E^{-2.13}$ & $E^{-2.5}$\\
\hline
IC86-1 & 222.3 & 48    & 44.3      & 11 & 6.4   & 0 & 0.051     & 63\,243    & 58.3 & 331.3 \\
IC86-2 & 330.4 & 73    & 67.9      & 7  & 9.3   & 0 & 0.082     & 94\,614    & 91.2& 541.7\\
IC86-3 & 362.0 & 72    & 69.3      & 6  & 6.3   & 0 & 0.073     & 103\,036   & 104.9 & 621.7\\
IC86-4 & 369.1 & 88    & 70.7      & 6  & 6.6   & 0 & 0.074     & 104\,846   & 107.0 & 633.9 \\
IC86-5 & 364.3 & 57    & 60.5      & 10 & 5.6   & 1 & 0.061     & 94\,699    & 105.6 & 625.6 \\
\hline
Sum   & 1648.1 & 338   & 312.7     & 40 & 34.2  & 1 & 0.341     & 460\,438  & 467.0 & 2754.2\\
\end{tabular}
\end{ruledtabular}
\begin{flushleft}
\small{$^a$ The subset of doublets which would trigger follow-up observations with the \emph{Swift} satellite. Until 2013-02-10 doublets with $\lambda<-8.8$ were forwarded to \emph{Swift} while later the threshold was changed to $-9.41$. This table includes alerts that could not be observed due to their proximity to the Sun or Moon.\\
$^b$ Here alerts consisting of exactly three neutrinos. Alerts with higher neutrino multiplicities were not detected and we expect only $(5.4 \pm 0.7)\times10^{-4}$ quadruplets from background over the complete livetime. The total number of background multiplets deviates from the number presented in \href{https://arxiv.org/abs/1702.06131}{Aartsen et al. (2016).}, because here we did not use data collected before Sept. 2011, but we did include times when the follow-up program was not running in realtime ($\sim3$\% of the data).\\
$^c$ Expected number of astrophysical track-like events obtained using simulated neutrino events which follow the measured spectra in the energy range from $100$\,GeV to $10$\,PeV. Events above 10\,TeV (true neutrino energy) make up 30\% (66\%) of the events for the $E^{-2.5}$ ($E^{-2.13}$) spectrum.}
\end{flushleft}
\end{table*}

\begin{table*}[p]
\caption{\label{tab:assumptions}Impact of assumptions on limits.}
\begin{ruledtabular}
\begin{tabular}{l l c c}
quantity & assumption & change in \# of multiplets & impact on limits\\
\hline
redshift distribution\,$^a$ & CCSNe (\href{https://arxiv.org/abs/1403.0007}{Madau \& Dickinson 2014}) & 1 & 1 \\
& CCSNe (\href{https://arxiv.org/abs/1509.06574}{Strolger et al. 2015} & 1.1 & 0.92 \\
& GRBs (\href{https://arxiv.org/abs/0912.0709}{Wanderman \& Piran 2010}) & 1.01 & 0.99 \\
& short GRBs (\href{https://arxiv.org/abs/1405.5878}{Wanderman \& Piran 2015}) & 1.2 & 0.87 \\
& no evolution & 4.1 & 0.37 \\
\hline
luminosity function\,$^b$ & CCSN-like (lognormal dist. with $\sigma=0.4$) & 1 & 1 \\
& GRB-like (\href{https://arxiv.org/abs/0912.0709}{Wanderman \& Piran 2010}) & 27 & 0.10 \\ 
& standard candle sources & 0.07 & 5.7 \\
\hline
source durations $t_{90}$ & GRB-like (\href{http://swift.gsfc.nasa.gov/archive/grb_table/}{\emph{Swift} GRB catalog}) & 1 & 1\\
& $\ll100\,$s$\,^c$ & 1.08 & 0.95 \\
& 200\,s & 0.62 & 1.3 \\
& 1000\,s & 0.08 & 5.3 \\
\hline
angular resolution & IceCube's optical follow-up program & 1 & 1\\
& $\ll 3.5^\circ\,^c$ & 1.27 & 0.87 \\
\hline
detector efficiency & including systematic uncertainties & 1 & 1\\
& without systematic uncertainties & 1.2 & 0.9 \\
\end{tabular}
\end{ruledtabular}
\begin{flushleft}
\small{The CCSNe-like population is here considered the standard scenario and the two last columns show by which factor the number of multiplets and the upper limit on the source energy change when variing the assumptions. The changes were calculated for the event selection of the 2013 -- 2015 season, an $E^{-2.5}$ spectrum and a source rate of $6.8\times 10^{-7}\,\text{Mpc}^{-3}\,\text{yr}^{-1}$ corresponding to $1\%$ of the CCSN rate.\\
$^a$ For a population of faint sources, the expected number of multiplets is determined by the fraction of neutrinos emitted by very nearby sources which is similar for the four upper redshift distributions. For the GRB rate ($4.2\times10^{-10}\,\text{Mpc}^{-3}\,\text{yr}^{-1}$) more distant transients become detectable such that redshift distribution has a different impact. Compared to the standard scenario the limits change by factors of 0.80, 1.12, 0.78 and 0.70 respectively for the four alternative redshift distributions.\\
$^b$ The large impact of the luminosity function is mostly caused by quoting the limit on the median bright source in the population. The same median source energy corresponds to $4.9$ times more neutrinos for the GRB-like population and to $3.8$ times fewer events for the standard candle scenario. If the source luminosity distributions would be aligned by their mean rather than the median each population would produce the same number of detected events. The limits on the mean source energy hence only deviate by a factor of 0.58 and 1.4 for the GRB-like and standard candle population.
$^c$ The ``$\ll$'' sign indicates that no losses occur due to the 100\,s or $3.5^\circ$ cut.}
\end{flushleft}
\end{table*}


\begin{figure*}[tb]
\begin{center}
\includegraphics[width=0.8\textwidth]{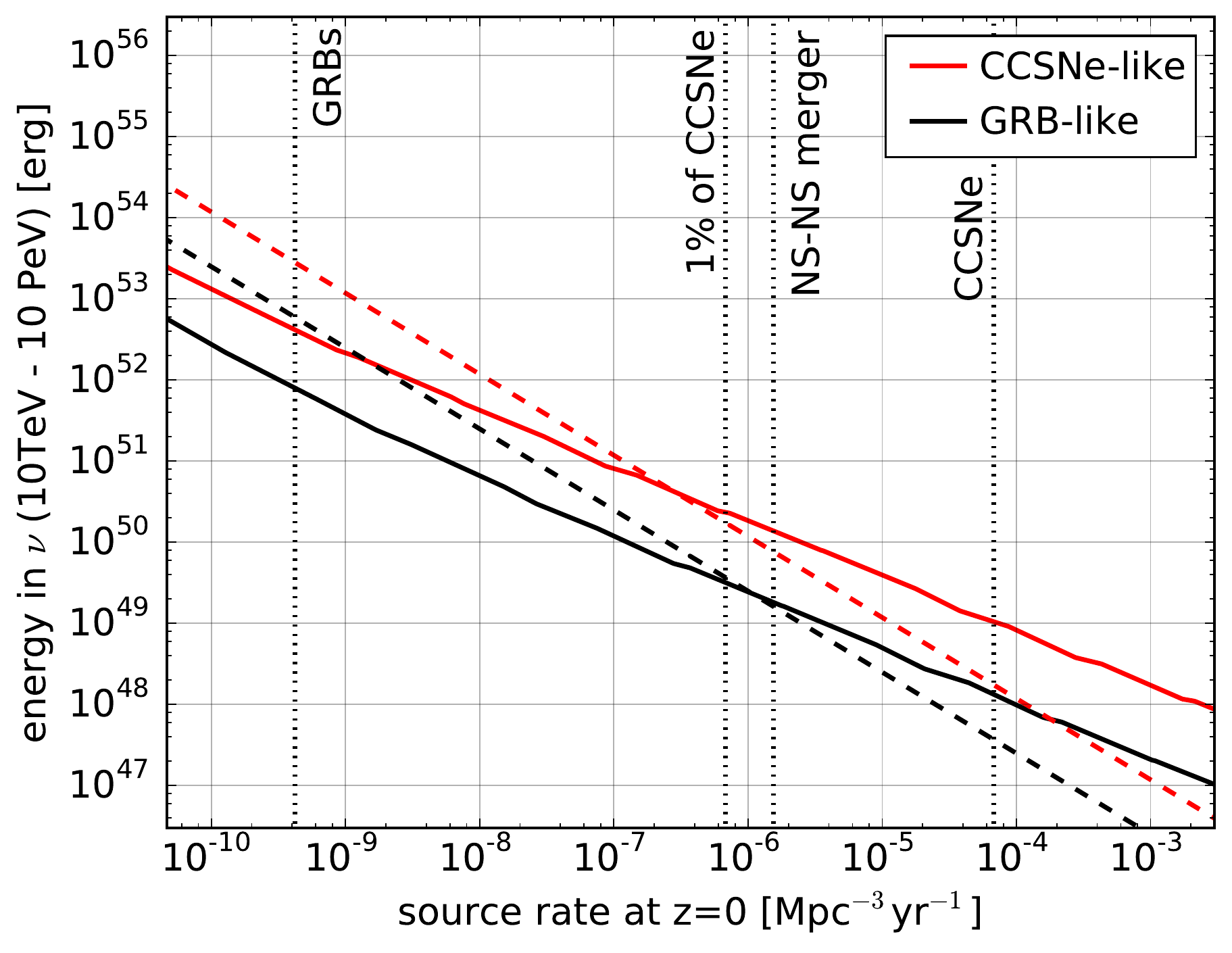}
\end{center}
\caption{\label{fig:limit_10tev} Limit on the energy released in neutrinos (all flavors) for the median source in the simulation. The limit was calculated as in Fig.~2 in the paper, except that we here only use neutrino events with a true energy between 10\,TeV and 10\,PeV. The result (solid lines) is the same for the $E^{-2.13}$ and $E^{-2.5}$ spectrum. The dashed diagonal lines show the source energy that would saturate the complete astrophysical flux for the $E^{-2.5}$ spectrum. The corresponding line for the $E^{-2.13}$ is lower by a factor of 3.}
\end{figure*}

\end{document}